\documentclass[prl,twocolumn,floatfix,showpacs]{revtex4}
\usepackage{graphicx}
\usepackage{amsmath}
\usepackage{amssymb}

\begin{document}
\title{Topological Kondo effect with Majorana fermions}
\author{B. B\'eri and N. R. Cooper}
\affiliation{TCM Group, Cavendish Laboratory, University of Cambridge, J.~J.~Thomson Ave., Cambridge CB3~0HE, UK}
\date{June 2012}
\begin{abstract}
The Kondo effect is a striking consequence of the coupling of itinerant
electrons to a quantum spin with degenerate energy levels. While degeneracies are commonly thought to arise from symmetries or fine-tuning of parameters, the recent emergence of Majorana fermions has brought to the fore an entirely different possibility: a  ``topological degeneracy" which arises from the nonlocal character of Majorana fermions. Here we show that nonlocal quantum spins formed from these degrees of freedom give rise to a novel ``topological Kondo effect". This leads to a robust non-Fermi liquid behavior, known to be difficult to achieve in the conventional Kondo context. Focusing on mesoscopic superconductor devices, we predict several unique transport signatures of this Kondo effect, which would demonstrate the non-local
quantum dynamics of Majorana fermions, and validate their potential for
topological quantum computation.
\end{abstract}
\pacs{73.23.-b,74.78.Na,72.10.Fk,03.67.Lx}
\maketitle

Traditionally, the Kondo effect arises when conduction electrons couple to  a
  confined region with a spin-degenerate ground state\cite{hewson1997kondo,pustilnik2004kondo}. More intricate
  scenarios, combining spin with other degeneracies, 
  can lead to exotic, non-Fermi liquid (NFL) behavior\cite{NozBlan,Mat95,CoxZaw,Oreg03,potok2007observation}. These
 degeneracies, however, require fine tuning of
  parameters, rendering such exotic physics quite fragile. Recent developments have shown that condensed
  matter systems can display another, much more robust, degeneracy
  called ``topological degeneracy''\cite{kitaev2003fault,nayak2008non}. This can arise
from the appearance of localized Majorana fermions in
  certain superconductor structures\cite{BeeMajrev,AliMajrev}.

The possibility of realizing Majorana fermions using superconductors has transformed an elusive notion of high-energy physics into  a tangible excitation in electronic materials\cite{wilczek2009majorana}. Methods for creating 
them in a controlled manner have been proposed, building on such simple ingredients as s-wave superconductors and spin-orbit coupling\cite{FuKane08,sau2010generic,alicea2010majorana,oreg2010helical}. This 
has led to the recent experimental observation\cite{Mourik25052012} of  conductance signatures indicating localized Majorana modes\cite{LawMaj,Flencond,Saucond,wimmer2011quantum}.
A key feature, not yet addressed experimentally, is that pairs of Majoranas can nonlocally encode zero energy fermions, which span a multidimensional ground state subspace. The degeneracy of the ground state is topological: it is ensured, up to exponentially small corrections, provided the Majoranas do not overlap. The resulting nonlocal zero-energy degrees of freedom form the topological qubits that underlie proposed schemes for fault tolerant quantum computation\cite{kitaev2003fault,nayak2008non,BeeMajrev}. Finding ``smoking gun" signatures of their quantum dynamics is an urgent issue.

\begin{figure}
\includegraphics[width=0.9\columnwidth]{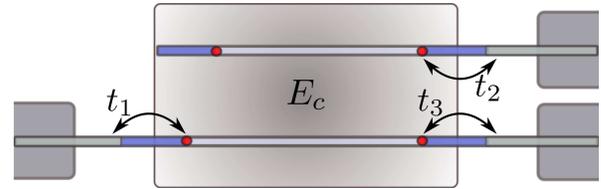}
\caption{Minimal setup for the topological Kondo effect. There are $M_\text{tot}=4$ Majorana fermions (red dots), $M=3$ of which are coupled to conduction electrons. The figure illustrates the realization based on semiconductor nanowires (horizontal bars). The wires are deposited on top of a superconductor (central rectangle). Nearby gates (not shown) put the central segment of the wires in a topological superconducting phase, while the adjacent segments are depleted, forming a tunnel barrier. The outermost segments host conduction electrons, and can be contacted to normal metal electrodes (outer rectangles).}
\label{fig:setup}
\end{figure}

The Kondo effect provides a central paradigm leading to observable consequences   of quantum dynamics within a degenerate ground state, but the possibility that the degeneracy has a topological origin has not previously been considered.
We will show that topological degeneracy can be a source of
novel exotic Kondo effects and  NFL behavior that is
  highly robust.  We predict that this ``topological Kondo
  effect'' leads to striking signatures in simple transport
  measurements on mesoscopic superconductor structures that support
  Majorana fermions. Such measurements can be used to give clear
  evidence for the quantum dynamics of the non-local qubits 
which form the basis for the proposed uses of
Majorana fermions in fault tolerant quantum computation.

We consider a setup consisting of a superconducting island, supporting  $M_\text{tot}$ localized Majorana modes, $M$ of which are coupled to spinless conduction electrons. The conduction electrons occupy $M$ single mode quantum wires (leads). As we explain below, the topological Kondo effect requires $M_\text{tot}\geq 4$, $M\geq 3$. 
The simplest configuration with minimal $M_\text{tot}$ and $M$ is shown in Fig.~\ref{fig:setup}.
 There can be several realizations, e.g. using superconducting heterostructures based on topological insulators\cite{FuKane08}, or semiconductor structures\cite{sau2010generic,alicea2010majorana,oreg2010helical} as in the nanowire setup in the experiment of Mourik {\it et al.}\cite{Mourik25052012}. We take the superconductor to be of mesoscopic size, connected to ground by a capacitor. It was noted in Ref.~\onlinecite{Futelep} that the charging energy $E_c$, which becomes relevant for superconductors in this regime, can play an important role related to Majorana fermions. It also has a key part in our considerations. It contributes to the Hamiltonian by a term
\begin{equation}
H_{c}(N)=E_{c}\left(N-\frac{q}{e}\right)^{2},\end{equation} 
where $N$ is the number of electrons on the island, and $q$ is a background charge determined by the voltage across the capacitor. The $M_\text{tot}$ Majorana modes correspond to $M_\text{tot}/2$ zero energy fermionic modes ($M_\text{tot}$ is always even). The parity of the total occupation number of these modes is tied to the parity of $N$.  Therefore, in each $N$ sector we have a $2^{M_\text{tot}/2-1}$-fold ground state degeneracy, which immediately shows why $M_\text{tot}\geq 4$ is required. The excited states above the ground state manifold are separated by a gap $\Delta$. For the realizations mentioned above, $\Delta\sim E_c \sim 0.5-1$K is a reasonable estimate\cite{Futelep,Sauexpt}

Working with temperatures and voltages $T,V\ll \Delta,E_c$, for weak lead-island coupling, the low energy physics is dominated by virtual transitions connecting the lowest energy ground state manifold of charge $eN$ to ground states with $N\pm1$ electrons. 
This physics is captured by the effective Hamiltonian  
\begin{equation}
H_{\text{eff}}=\sum_{i\neq j}\lambda^{+}_{ij}\gamma_{j}\gamma_{i}\psi_{i}^{\dagger}\psi_{j}-\sum_i\lambda^{-}_{ii}\psi_{i}^{\dagger}\psi_{i},
\label{eq:heffgam}
\end{equation}
where we have introduced the constants $\lambda^\pm_{ij}=\left(\frac{1}{U_{+}}\pm\frac{1}{U_{-}}\right)t_{i}t_{j}$, with $U_{\pm}=H_{c}(N\pm1)-H_{c}(N)$ and tunneling amplitudes $t_i$ (which can always be chosen positive).  Eq.~\eqref{eq:heffgam} is obtained by  a Schrieffer-Wolff transformation\cite{hewson1997kondo}, implementing the leading order perturbation theory in the lead-island couplings. 
The full Hamiltonian is  $H=H_\text{lead}+H_\text{eff}$, where the first term is the Hamiltonian of the conduction electrons, which we assume to be noninteracting.

To illuminate how the Kondo problem emerges,  let us focus on the first term in Eq.~\eqref{eq:heffgam} and consider the setup of Fig.~\ref{fig:setup} with $M=3$  coupled Majoranas. It is known (see e.g. Ref.~\onlinecite{BeeMajrev}) 
that the three $\gamma_i$ realize a spin-1/2 object 
\begin{equation}
\sigma_1=-i\gamma_2\gamma_3,\ \sigma_2=i\gamma_1\gamma_3,\ \sigma_3=-i\gamma_1\gamma_2.
\label{eq:sigmas}\end{equation}
Coupling this to the three species in the leads
suggests a Kondo problem of a spin-$\frac{1}{2}$ impurity with spin-1 conduction electrons\cite{FabGog,SenKim96}. Indeed, we have a Kondo term
\begin{equation}
\sum_{i\neq j}\lambda^{+}_{ij}\gamma_{j}\gamma_{i}\psi_{i}^{\dagger}\psi_{j}=\frac{1}{2}\sum_\alpha \lambda_\alpha \sigma_\alpha J_\alpha\label{eq:majkon}
\end{equation} 
with $\lambda_\alpha=\sum_{ab}|\varepsilon_{\alpha ab}|\lambda^{+}_{ab}$, where the conduction electrons enter through the spin-1 object  $J_\alpha\!=\!i\sum_{ab}\varepsilon_{\alpha ba}\psi^\dagger_a \psi_b$. Remarkably, the spin structure of $J_\alpha$ is distributed nonlocally to spatially separate leads; this will result in distinct transport signatures.

The Kondo term is nontrivial if the impurity acts as a quantum spin, as opposed to a classical Ising variable. This requires coupling to at least two of the $\sigma_\alpha$. This needs three $\gamma_j$, showing why $M=3$ is the minimal case. 
The same $\sigma_\alpha$ are the Pauli matrices acting on the topological qubit\cite{BeeMajrev}.  The topological Kondo effect reveals the quantum spin nature of this object, thereby detecting the quantum qubit dynamics. A ``smoking gun" signature of this is already clear: this Kondo effect should disappear if any one of the three leads is decoupled.

To see how the Kondo effect shows up, we begin with a renormalization group (RG)  analysis of the minimal setup of Fig.~\ref{fig:setup}. 
Our considerations also apply for $M_\text{tot}>4$, allowing for stray Majoranas not coupled to the leads. The presence of these modes is akin to the presence of uncoupled spins not participating in the Kondo effect.

In terms of bare parameters, $H_\text{eff}$ enters as  a weak perturbation. We obtain the RG flow in this weak coupling regime using the poor man's scaling procedure, giving
\begin{equation}
\frac{d\lambda_1}{dl}=\rho \lambda_2\lambda_3,\ \ \text{cycl. perm.},\label{eq:kondoweak}
\end{equation}
where $\rho$ is the density of states of the leads at the Fermi energy. 
 The couplings $\lambda^-_{ij}$, similar to the potential scattering terms in the  Kondo context, do not renormalize. These are the usual weak coupling RG equations of the Kondo problem, but it should be kept in mind that  $\lambda_\alpha$ now characterize nonlocal charge transfers  between different leads. As both $U_\pm,\ t_i>0$, the bare Kondo coupling is antiferromagnetic. Typically $t_i$ will not have the same value, which translates into an exchange anisotropy in the Kondo language. Under Eq.~\eqref{eq:kondoweak} the couplings increase, while  $\lambda_\alpha^2-\lambda_\beta^2$ remain constant. The flow is therefore towards an isotropic coupling,  $\lambda_\alpha/\lambda_\beta\rightarrow 1$.  This conventional result in the Kondo context translates into something remarkable for our setup: 
 a tendency towards a
 threefold symmetry with respect to relabeling the leads $j\rightarrow j\!+\!1 (\text{mod} 3)$. The overall behavior of the couplings is characterized by an inverse logarithmic growth
\begin{equation}
\lambda_\alpha(\Lambda)\sim\frac{1}{\ln(\Lambda/T_{\rm K})},
\label{eq:invlog}
\end{equation}
where we have introduced the Kondo temperature $T_{\rm K}$, and the renormalized high-energy cutoff $\Lambda=\Lambda_0 e^{-l}\sim E_c e^{-l}$. Denoting by $\bar{\lambda}$ a typical bare value of the $\lambda_\alpha$-s, one has $T_{\rm K}\sim E_c e^{-1/\rho\bar{\lambda}}$.
The factors entering $T_{\rm K}$ are the same as for conventional Kondo arrangements, which implies that considering Kondo temperatures anywhere in the range $0<T_{\rm K}\lesssim 0.1$K is reasonable\cite{potok2007observation}.

Upon approaching $\Lambda\sim T_{\rm K}$, the couplings cease to be small, and the perturbative RG has to be replaced by a nonperturbative analysis. A powerful route is provided by the conformal field theory (CFT) method of Affleck and Ludwig\cite{affleck1990current,affleck1991kondo,affleck1991critical}, which we applied to our problem\cite{app}. (This complements Ref.~\onlinecite{FabGog}, where the  Kondo problem of spin-$1$ electrons was studied using abelian bosonization assuming an axial symmetry not present in our case.) 
We find that the flow is towards an intermediate coupling fixed point with NFL behavior which is robust. In the vicinity of the fixed point the scale dependence is due to dimension $4/3$ irrelevant operators $\mathcal{O}_\alpha$. The NFL behavior stems from these: $\mathcal{O}_\alpha$  cannot be constructed out of ordinary fermions, as fermions can give only halfinteger dimensions. In addition, marginal operators, corresponding to the original $\lambda^-_{ii}$ terms will also be present, but  their scale dependence (which is also through $\mathcal{O}_\alpha$) can be neglected.  
It is by itself remarkable that our setup, with simple noninteracting leads, allows the appearance of NFL behaviour without fine tuning of the couplings. This is unlike conventional Kondo variants leading to NFL physics, where one has to fine tune at least one coupling\cite{NozBlan,Mat95,CoxZaw,Oreg03,potok2007observation}, or introduce leads which are themselves NFLs\cite{FabGog2c,FieteNayak,Law2c}.

The weak coupling flow and the knowledge of the nature of the intermediate coupling fixed point can be applied to deduce the behavior of various experimentally relevant quantities. Given that the  first signatures of localized Majorana modes were obtained by conductance measurements, we focus on the conductance $G_{kl}$ between leads $k$ and $l$. For simplicity, we work in the linear response regime and focus on the temperature dependence of $G_{kl}$. (The results also apply to the nonlinear differential conductance in the opposite, $T\ll V$ case, upon replacing $T$ by $V$ in the expressions.) The key findings,  Eqs.~\eqref{eq:Gwc} and \eqref{eq:NFL} below,  are summarized in Fig.~\ref{fig:GTdep}. We emphasize that all features come with an extra, ``smoking-gun" signature: the signs of the Kondo effect disappear from $G_{kl}$ upon decoupling the third ($j\neq k,l$) lead. 

In the weak coupling regime, the only term in the Hamiltonian \eqref{eq:heffgam}  which  transfers charge between the leads is the Kondo coupling. 
The conductance is therefore $G_{kl}\sim (\lambda^{+}_{kl})^2$ to leading order in $\lambda^{+}_{kl}$. Combining this with the scaling \eqref{eq:invlog}, we find the behavior for $T_{\rm K}\ll T$, 
\begin{equation}
G_{kl}\sim\frac{1}{\ln^{2}(T/T_{\rm K})}.\label{eq:Gwc}\end{equation}
Observing such an inverse logarithmic increase would be a qualitative signature of the topological Kondo effect.  
 Through $G_{kl}\sim (\lambda^{+}_{kl})^2$ the conductances also provide a direct measure of the degree of anisotropy of the Kondo coupling. As long as Eq.~\eqref{eq:Gwc} is valid, the scaling \eqref{eq:kondoweak} translates to the temperature independence of $G_{ij}-G_{kl}$ while $G_{ij}/G_{kl}\rightarrow 1$ as $T$ is lowered. Observing this tendency would be another qualitative signature.

At low temperatures,  $T\!\!\ll\!\! T_{\rm K}$, the inverse logarithmic increase crosses over to a power law convergence to the zero temperature limit. Suppressing the small, temperature independent corrections due to the marginal operators, we find\cite{app}
\begin{equation}
G_{kl}(T)=\frac{2e^2}{3h}+c_{kl}T^{2/3}\qquad  (k\neq l),\label{eq:NFL}
\end{equation}
where the temperature dependence is due to a second order correction in the irrelevant operators $\mathcal{O}_\alpha$ and $c_{kl}$ are nonuniversal coefficients. (Our simple scaling analysis does not tell the sign of $c_{kl}$, but we expect $G_{kl}$ to continue its monotonic increase implying $c_{kl}<0$.) The diagonal conductances follow from $G_{kl}$ through current conservation, and they approach $\frac{4e^2}{3h}$ as $T\rightarrow 0$. That this value exceeds the conductance quantum indicates the presence of Andreev reflection processes, allowing for holes, not only electrons, to be backscattered. Note that these are of different origin than in usual normal-superconducting systems, where the superconductor absorbs a Cooper pair in the process. Indeed, the charging energy forbids this in our case. Instead, our system realizes a strongly correlated ``Andreev reflection fixed point"\cite{Nayak99,Oshikawa06}, with the two electrons playing the role of Cooper pairs exiting through the leads. Detecting this enhanced conductance together with the $T^{2/3}$ dependence would be a clear signature of the NFL physics.

\begin{figure}
\includegraphics[width=0.9\columnwidth]{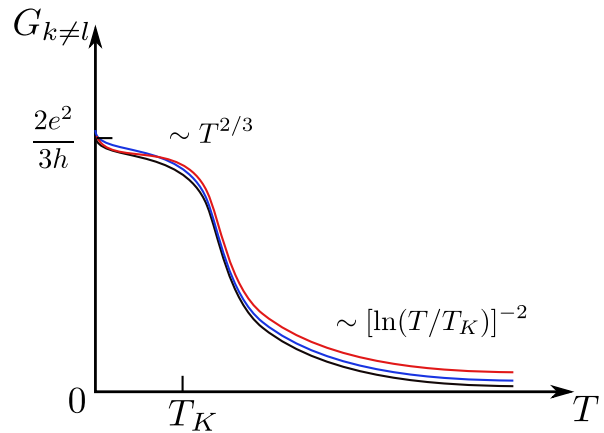}
\caption{The predicted signatures of the topological Kondo effect. The three curves represent the temperature dependence of the three offdiagonal conductances of the setup in Fig.~\ref{fig:setup}. In the generic case the curves cross each other, because their separation at high and low temperatures are due to different reasons: the anisotropy of the Kondo coupling for $T\gg T_{\rm K}$ and the marginal perturbations for $T\ll T_{\rm K}$.}
\label{fig:GTdep}
\end{figure}

Before concluding, we briefly discuss the generalization of our results to $M>3$. 
The $M$ Majorana operators generate a Clifford algebra\cite{wilczek2009majorana}.  This implements the spinor representation of the Lie algebra of the orthogonal group SO$(M)$\cite{zee2010quantum,fuchs2003symmetries}, with $i\gamma_j\gamma_k$ ($j<k$) representing the $jk$-th SO$(M)$ generator. These generalize  $\sigma_\alpha$ in Eq.~\eqref{eq:majkon}. $i\epsilon_{\alpha ab}$ generalizes to  $A_{ab}^{(jk)}=i\left(\delta_{ja}\delta_{kb}-\delta_{ka}\delta_{jb}\right)$,  the $jk$-th SO$(M)$ generator in the defining representation. For general $M<M_\text{tot}$ we thus have an  SO$(M)$ Kondo problem, with a spinor impurity and conduction electrons in the defining representation.
For $M=M_\text{tot}$ the impurity is in a half-spinor representation depending on the parity of $N$. When this is faithful, lead electrons again furnish the defining SO$(M)$ representation.
To the best of our knowledge, Kondo problems of this type did not appear in the literature so far. In particular, these problems are markedly different from the descriptions of the two-channel Kondo model related to orthogonal groups and/or Majorana fermions\cite{EK,ColemanKondo1,ColemanKondo2,maldacena1997majorana}. In addition to the apparent distinction that these works introduce Majoranas only for mathematical convenience, their models themselves are different from ours: they do not conserve charge\cite{ColemanKondo1,ColemanKondo2}, or have different group structure\cite{EK,maldacena1997majorana}.

We end by outlining some features of the general $M<M_\text{tot}$ case, assuming isotropic couplings $\lambda^+_{ij}=\lambda^+$, $\lambda^-_{ii}=\lambda_-$. (We expect that the results also hold for $M=M_\text{tot}$ with faithful half-spinor representations.) The scaling \eqref{eq:kondoweak} generalizes to 
\begin{equation}
\frac{d\lambda^+}{dl}=2\rho (M-2)(\lambda^+)^2,
\end{equation}
while $\lambda^-$ does not renormalize. This again implies an inverse logarithmic growth of $\lambda^+$ and the corresponding inverse log-square temperature dependence of the weak coupling conductance. At low temperatures, we expect NFL behavior, with a convergence to $G_{kl}=\frac{2e^2}{Mh}$ for $k\neq l$, obtained by generalizing the results\cite{Nayak99,Oshikawa06} for the Andreev reflection fixed point. 

In summary, we have shown that the topological degeneracy of
  Majorana fermions can lead to a new class of ``topological'' Kondo
  effects.  
  These effects are not only novel from a
  mathematical perspective, but have important and striking physical
  consequences for realistic experimental systems.  We have
  established the detailed properties for the simplest case (with $M=3$ leads coupled to  Majorana fermions) in which the topological degeneracy
  gives rise to a dynamical non-local quantum spin.  We have shown that
  this leads to a non-Fermi liquid behavior that is robust to
  perturbations, in contrast to the conventional Kondo context, where such behavior is known to be unstable. The resulting non-trivial power law dependences and the enhanced conductance due to strong correlations are all distinctive qualitative features, which come with a ``smoking gun" signature: they can be switched off at will by decoupling any one of the three leads. The physics we describe can   readily be explored in experiments on mesoscopic devices based on   superconducting structures using available technology.  These studies would provide a clear test of the expected non-local quantum
  dynamics of Majorana fermions: such a measurement would be a crucial
  step towards establishing the Majorana architecture for
  fault-tolerant quantum computation.

\acknowledgments{We acknowledge useful discussions with D. E. Logan and N. d'Ambrumenil. This work was supported by EPSRC Grant EP/F032773/1.}

\section*{Supplementary Material to "Topological Kondo effect with Majorana fermions"}
In this Supplementary Material, we briefly summarize the application of the conformal field theory (CFT) method of Affleck and Ludwig\cite{affleck1990current,affleck1991kondo,affleck1991critical} to our problem. 
A standard initial step in the CFT approach is to convert the fermion operators with left and right moving pieces $\psi_{Lj}$, $\psi_{Rj}$, defined for $x\geq 0$ in each lead, into a left moving field $\psi_{Lj}$, extended to $x<0$. 
 In terms of these, the full Hamiltonian reads 
\begin{multline}
 H=v\sum_j\int  \psi_{Lj}^\dagger (x)i\partial_x\psi_{Lj}(x)dx+g^{(c)} J^{(c)}(0)\\ + \sum_\alpha g^{(s)}_\alpha S_\alpha J^{(s)}_\alpha(0)+g^{(d)}_\alpha J^{(d)}_\alpha(0),
\label{eq:HKL}\end{multline}
 where we introduced the rescaled real coupling constants $g^{(j)}_\alpha$, with $g^{(s)}_\alpha>0$, in particular.  
(The precise form of the rescaling depends on the boundary condition for the conduction electrons, which we do not need to specify for our purposes.)
The first term is $H_\text{lead}$ in its low energy form with a linearized spectrum characterized by a Fermi velocity $v$.  The remaining terms come from $H_\text{eff}$. They involve the densities
\begin{equation}
J^{(c)}(x)=\vec{\psi}^{\dagger}(x)\cdot\vec\psi(x),\   J^{(s)}_\alpha(x)=\vec\psi^\dagger(x) T_\alpha\vec\psi(x),
\end{equation}
and $J^{(d)}_\alpha(x)=\vec\psi^\dagger(x) D_\alpha\vec\psi(x)$, where $D_\alpha$ are diagonal traceless real matrices. Here we used a matrix-vector notation and dropped the $L$ index for brevity. 
 Central to the CFT method is that $H_\text{lead}$ can also be expressed entirely in terms of the charge and spin densities $J^{(c,s)}$, allowing for the problem to be recast in the language of the corresponding Kac-Moody algebras. The CFT description works in the vicinity of an isotropic fixed point (FP). If this is stable, our weak coupling RG makes it plausible to interpret it as the Kondo FP to which the system flows upon leaving the weak coupling regime. 

The perturbations around the FP are organized into conformal families. For spin-1 electrons, the density $J^{(s)}$ satisfies an SU$(2)_4$ Kac-Moody algebra\cite{FabGog}, hence the relevant conformal families are built from products from those of a SU$(2)_4$ and a U$(1)$ theories, the latter accounting for the charge sector. Focusing only on charge conserving perturbations, the families  are uniquely labeled by the spin of the operator of the lowest scaling dimension, the spin-$j$ primary with  $j=0,\frac{1}{2},\ldots,2$. The allowed perturbations have to satisfy two criteria: (i) they have to respect time-reversal (TR) invariance and (ii) they have to be in a family which can be obtained from the ones of the unperturbed, free fermion theory through a double fusion with the spin-$\frac{1}{2}$ primary.

The first criterion holds in spite of the fact that the original Hamiltonian breaks TR invariance (due to the magnetic fields needed to realize Majorana modes). The reason is that the emerging Kondo problem is invariant under a "time-reversal" implemented by complex conjugation in the $\psi$ sector, sending $J_s\!\rightarrow\! -J_s$, accompanied by $S\!\rightarrow\! -S$ in the impurity sector. The second condition is the consequence of the fusion hypothesis of Affleck and Ludwig\cite{affleck1990current,affleck1991kondo,affleck1991critical}, a key element of the CFT method. 

The charge conserving operators in the free fermion  theory have to contain the same number of creation and annihilation operators. The primary operators which can be constructed with this constraint are the $j=0$ primary (the identity) and the $j=2$ primary, realized by the five field multiplet
\begin{equation}
J^{(2)}_\alpha(x)=\vec\psi^\dagger(x) C_\alpha\vec\psi(x),
\end{equation}
with a real symmetric traceless matrix $C_\alpha$.  The allowed perturbations around the FP come from the families obtained after the SU$(2)_4$ double fusion, which sends
\begin{equation}
j=0\rightarrow j=0,1,\quad j=2\rightarrow j=1,2.\end{equation}
Note that now we have the $j=1$ primary, which is not in the free fermion spectrum. Anything deriving from the corresponding family will lead to a NFL behavior. 

We now have to check which are the leading TR invariant perturbations, i.e. the ones with the smallest scaling dimension. The spin-$j$ primary operators themselves are TR invariant for $j=0,2$, and so is the first subleading perturbation in the $j=0$ family, the charge density $J^{(c)}$. While the the $j=0$ primary is simply the identity, the other two operators are nontrivial perturbations, both with scaling dimension $\Delta^{(0,2)}=1$. The $j=1$ primary $\vec{\phi}$ flips under TR, but its first level descendants (constituting the first subleading level), $J^\alpha_s \phi^\beta$ respect TR invariance. Their scaling dimension is $\Delta^{(1)}=\frac{4}{3}$. These are the leading  
allowed perturbations.  The first order RG equations around the FP for the corresponding coupling constants $K^{(j)}$ (suppressing the multiplet index $\alpha$) are $\frac{dK^{(j)}}{dl}=[1-\Delta^{(j)}]K^{(j)}$, which shows that $K^{(0,2)}$ are marginal, while $K^{(1)}$ are the leading irrelevant couplings. 

The possible presence of the marginal $K^{(0,2)}$ couplings is not surprising: the corresponding terms are already there in the bare Hamiltonian through $g^{(c,d)}$. As $K^{(0,2)}$ are in the free fermion spectrum and the SU$(2)_4$ fusion rules dictate that the corresponding families can fuse only into each other, they are expected to be scale independent for  $K^{(1)}=0$ even if higher order terms are included in the RG. For $K^{(0)}$, this remains true also for $K^{(1)}\neq 0$, as $K^{(1)}$ act only in the spin sector.
The scale independence of $K^{(2)}$ for  $K^{(1)}=0$ implies that for  $K^{(1)}\neq 0$ all terms in the RG equations involve $K^{(1)}$, which means that $K^{(2)}$ essentially stay constant also in this case; their only possible $l$ dependence being a decaying small amplitude correction through the $l$ dependence of $K^{(1)}$. We thus conclude that the CFT FP is stable, and we can consistently identify it with the Kondo FP. We also find that all the scale dependence is through $K^{(1)}$; the operators denoted by $\mathcal{O}_\alpha$ in the main text correspond to these couplings. As noted in the main text, the presence of the operators $\mathcal{O}_\alpha$ leads to non-Fermi liquid behavior. This implies that the fixed point is at intermediate coupling: one can show that the infinite coupling problem corresponds to an effective noninteracting Hamiltonian, i.e. to Fermi liquid physics.

The CFT method also allows us to obtain predictions for the low temperature behavior of the conductance. The key quantities are the correlation functions of the currents $I_{Xj}(x)=\psi_{Xj}^\dagger(x)\psi_{Xj}(x)$, where $X=L,R$ labels left and right movers (defined for $x\geq 0$) and $j=1,2,3$ labels the leads. A general CFT argument of Ref.~\onlinecite{Rahmani10} shows that the FP conductance $G^*_{jk}$ between leads $j$ and $k$  ($j\neq k$) requires only $\langle I_{Lj}I_{Rk}\rangle$. Due to the effective time-reversal symmetry we have   $\langle I_{Lj}I_{Rk}\rangle=\langle I_{Lk}I_{Rj}\rangle$. Combining this with the isotropy of the couplings we get $\langle I_{Lj}I_{Rj}\rangle=\langle I_{L1}I_{R1}\rangle$ and  $\langle I_{Lj}I_{Rk}\rangle=\langle I_{L1}I_{R2}\rangle$, thus 
\begin{equation}
\langle J_{L}^{(c)}J_{R}^{(c)}\rangle=3\langle I_{L1}J_{R1}\rangle+6\langle I_{L1}I_{R2}\rangle
\end{equation}
and
\begin{equation}
\langle J_{L}^{(d)}J_{R}^{(d)}\rangle=2\langle I_{L1}J_{R1}\rangle-2\langle I_{L1}I_{R2}\rangle
\end{equation}
where we used $D_1=\text{diag}(1,-1,0)$ in $J^{(d)}$. (Our considerations give the same result for any choice of a traceless diagonal matrix.) 
The correlation functions $\langle J_{L}^{(c)}J_{R}^{(c)}\rangle$, $\langle J_{L}^{(d)}J_{R}^{(d)}\rangle$ are two point functions of primary fields, which acquire only an overall factor due to the Kondo effect\cite{ludwig1991exact} compared to their free fermion values. Evaluating this factor following Ref.~\onlinecite{ludwig1991exact} we find that it is unity for $\langle J_{L}^{(c)}J_{R}^{(c)}\rangle$, and minus one for $\langle J_{L}^{(d)}J_{R}^{(d)}\rangle$. This means that the correlation function $\langle I_{L1}I_{R2}\rangle$ itself acquires a factor $A_{12}=2/3$ compared to its free fermion value. We can invoke the result of Ref.~\onlinecite{Rahmani10} 
\begin{equation}
G^*_{12}=A_{12}\frac{e^2}{h},
\end{equation}
to find that the FP conductance is $G^*_{12}=\frac{2e^2}{3h}$. 
This result can be further supported by using abelian bosonization to map our original $H_\text{eff}$, in the isotropic limit, to tri-junction problems studied in Refs.~\onlinecite{Nayak99,Oshikawa06}. A key point to note is that  because the objects $\gamma_i\psi_i$ commute with each other instead of anticommuting, they can be bosonized without Klein factors, mapping our problem onto the "auxiliary problems" studied in those papers. These problems, which are unphysical illustrations in the original tri-junction context because of the absence of Klein factors, are known to lead to a strongly correlated Andreev reflection FP with $G^*_{12}=\frac{2e^2}{3h}$, even for noninteracting leads. Our setup provides a physical realization of this phenomenon. 

Finally we would like to show that the first order contributions to the conductance from the perturbations around the FP vanish. Away from the FP, our previous symmetry considerations do not apply, and the currents $I_{Xj}$ take the general form $I_{Xj}=a J^{(c)}+\sum_\alpha b_\alpha J^{(d)}_\alpha$. The first order contribution to the conductance originates from the three point functions
\begin{equation}
\langle J_L^{(p)}J_R^{(q)}\mathcal{P}\rangle,
\end{equation}
where $J^{(p,q)}$ can be one of the currents $J^{(c,d)}$ and $\mathcal{P}$ can be $J^{(c,d)}$ or $\mathcal{O}$, the leading allowed perturbations. These three point functions always vanish: when all three fields are in the spin sector they vanish due to the SU$(2)_4$ fusion rule $2\times 2\rightarrow 0$; when all three fields are $J^{(c)}$, we have a three point function of a free boson, which again vanishes; and in all the other cases they factorize and vanish because of the vanishing of one point functions of primary fields.   
These arguments, however, permit second order corrections which lead to the predicted $T^{2/3}$ temperature dependence.


\begin{thebibliography}{42}
\expandafter\ifx\csname natexlab\endcsname\relax\def\natexlab#1{#1}\fi
\expandafter\ifx\csname bibnamefont\endcsname\relax
  \def\bibnamefont#1{#1}\fi
\expandafter\ifx\csname bibfnamefont\endcsname\relax
  \def\bibfnamefont#1{#1}\fi
\expandafter\ifx\csname citenamefont\endcsname\relax
  \def\citenamefont#1{#1}\fi
\expandafter\ifx\csname url\endcsname\relax
  \def\url#1{\texttt{#1}}\fi
\expandafter\ifx\csname urlprefix\endcsname\relax\def\urlprefix{URL }\fi
\providecommand{\bibinfo}[2]{#2}
\providecommand{\eprint}[2][]{\url{#2}}

\bibitem[{\citenamefont{Hewson}(1997)}]{hewson1997kondo}
\bibinfo{author}{\bibfnamefont{A.}~\bibnamefont{Hewson}},
  \emph{\bibinfo{title}{The Kondo problem to heavy fermions}}
  (\bibinfo{publisher}{Cambridge University Press}, \bibinfo{year}{1997}).

\bibitem[{\citenamefont{Pustilnik and Glazman}(2004)}]{pustilnik2004kondo}
\bibinfo{author}{\bibfnamefont{M.}~\bibnamefont{Pustilnik}} \bibnamefont{and}
  \bibinfo{author}{\bibfnamefont{L.}~\bibnamefont{Glazman}},
  \bibinfo{journal}{J. Phys.: Condens. Matter} \textbf{\bibinfo{volume}{16}},
  \bibinfo{pages}{R513} (\bibinfo{year}{2004}).

\bibitem[{\citenamefont{Nozi{\`e}res and Blandin}(1980)}]{NozBlan}
\bibinfo{author}{\bibfnamefont{P.}~\bibnamefont{Nozi{\`e}res}}
  \bibnamefont{and} \bibinfo{author}{\bibfnamefont{A.}~\bibnamefont{Blandin}},
  \bibinfo{journal}{J.Phys.} \textbf{\bibinfo{volume}{41}},
  \bibinfo{pages}{193} (\bibinfo{year}{1980}).

\bibitem[{\citenamefont{Matveev}(1995)}]{Mat95}
\bibinfo{author}{\bibfnamefont{K.~A.} \bibnamefont{Matveev}},
  \bibinfo{journal}{Phys. Rev. B} \textbf{\bibinfo{volume}{51}},
  \bibinfo{pages}{1743} (\bibinfo{year}{1995}).

\bibitem[{\citenamefont{Cox and Zawadowski}(1998)}]{CoxZaw}
\bibinfo{author}{\bibfnamefont{D.}~\bibnamefont{Cox}} \bibnamefont{and}
  \bibinfo{author}{\bibfnamefont{A.}~\bibnamefont{Zawadowski}},
  \bibinfo{journal}{Adv. Phys.} \textbf{\bibinfo{volume}{47}},
  \bibinfo{pages}{599} (\bibinfo{year}{1998}).

\bibitem[{\citenamefont{Oreg and Goldhaber-Gordon}(2003)}]{Oreg03}
\bibinfo{author}{\bibfnamefont{Y.}~\bibnamefont{Oreg}} \bibnamefont{and}
  \bibinfo{author}{\bibfnamefont{D.}~\bibnamefont{Goldhaber-Gordon}},
  \bibinfo{journal}{\prl} \textbf{\bibinfo{volume}{90}},
  \bibinfo{pages}{136602} (\bibinfo{year}{2003}).

\bibitem[{\citenamefont{Potok et~al.}(2007)\citenamefont{Potok, Rau, Shtrikman,
  Oreg, and Goldhaber-Gordon}}]{potok2007observation}
\bibinfo{author}{\bibfnamefont{R.~M.} \bibnamefont{Potok}},
  \bibinfo{author}{\bibfnamefont{I.~G.} \bibnamefont{Rau}},
  \bibinfo{author}{\bibfnamefont{H.}~\bibnamefont{Shtrikman}},
  \bibinfo{author}{\bibfnamefont{Y.}~\bibnamefont{Oreg}}, \bibnamefont{and}
  \bibinfo{author}{\bibfnamefont{D.}~\bibnamefont{Goldhaber-Gordon}},
  \bibinfo{journal}{Nature} \textbf{\bibinfo{volume}{446}},
  \bibinfo{pages}{167} (\bibinfo{year}{2007}).

\bibitem[{\citenamefont{Kitaev}(2003)}]{kitaev2003fault}
\bibinfo{author}{\bibfnamefont{A.}~\bibnamefont{Kitaev}},
  \bibinfo{journal}{Annals of Physics} \textbf{\bibinfo{volume}{303}},
  \bibinfo{pages}{2} (\bibinfo{year}{2003}).

\bibitem[{\citenamefont{Nayak et~al.}(2008)\citenamefont{Nayak, Simon, Stern,
  Freedman, and Sarma}}]{nayak2008non}
\bibinfo{author}{\bibfnamefont{C.}~\bibnamefont{Nayak}},
  \bibinfo{author}{\bibfnamefont{S.}~\bibnamefont{Simon}},
  \bibinfo{author}{\bibfnamefont{A.}~\bibnamefont{Stern}},
  \bibinfo{author}{\bibfnamefont{M.}~\bibnamefont{Freedman}}, \bibnamefont{and}
  \bibinfo{author}{\bibfnamefont{S.}~\bibnamefont{Sarma}},
  \bibinfo{journal}{\rmp} \textbf{\bibinfo{volume}{80}}, \bibinfo{pages}{1083}
  (\bibinfo{year}{2008}).

\bibitem[{\citenamefont{Beenakker}()}]{BeeMajrev}
\bibinfo{author}{\bibfnamefont{C.~W.~J.} \bibnamefont{Beenakker}},
  \bibinfo{note}{arXiv:1112.1950}.

\bibitem[{\citenamefont{Alicea}()}]{AliMajrev}
\bibinfo{author}{\bibfnamefont{J.}~\bibnamefont{Alicea}},
  \bibinfo{note}{arXiv:1202.1293}.

\bibitem[{\citenamefont{Wilczek}(2009)}]{wilczek2009majorana}
\bibinfo{author}{\bibfnamefont{F.}~\bibnamefont{Wilczek}},
  \bibinfo{journal}{Nat. Phys.} \textbf{\bibinfo{volume}{5}},
  \bibinfo{pages}{614} (\bibinfo{year}{2009}).

\bibitem[{\citenamefont{Fu and Kane}(2008)}]{FuKane08}
\bibinfo{author}{\bibfnamefont{L.}~\bibnamefont{Fu}} \bibnamefont{and}
  \bibinfo{author}{\bibfnamefont{C.}~\bibnamefont{Kane}},
  \bibinfo{journal}{\prl} \textbf{\bibinfo{volume}{100}},
  \bibinfo{pages}{96407} (\bibinfo{year}{2008}).

\bibitem[{\citenamefont{Sau et~al.}(2010{\natexlab{a}})\citenamefont{Sau,
  Lutchyn, Tewari, and Das~Sarma}}]{sau2010generic}
\bibinfo{author}{\bibfnamefont{J.}~\bibnamefont{Sau}},
  \bibinfo{author}{\bibfnamefont{R.}~\bibnamefont{Lutchyn}},
  \bibinfo{author}{\bibfnamefont{S.}~\bibnamefont{Tewari}}, \bibnamefont{and}
  \bibinfo{author}{\bibfnamefont{S.}~\bibnamefont{Das~Sarma}},
  \bibinfo{journal}{\prl} \textbf{\bibinfo{volume}{104}},
  \bibinfo{pages}{40502} (\bibinfo{year}{2010}{\natexlab{a}}).

\bibitem[{\citenamefont{Alicea}(2010)}]{alicea2010majorana}
\bibinfo{author}{\bibfnamefont{J.}~\bibnamefont{Alicea}},
  \bibinfo{journal}{\prb} \textbf{\bibinfo{volume}{81}},
  \bibinfo{pages}{125318} (\bibinfo{year}{2010}).

\bibitem[{\citenamefont{Oreg et~al.}(2010)\citenamefont{Oreg, Refael, and von
  Oppen}}]{oreg2010helical}
\bibinfo{author}{\bibfnamefont{Y.}~\bibnamefont{Oreg}},
  \bibinfo{author}{\bibfnamefont{G.}~\bibnamefont{Refael}}, \bibnamefont{and}
  \bibinfo{author}{\bibfnamefont{F.}~\bibnamefont{von Oppen}},
  \bibinfo{journal}{\prl} \textbf{\bibinfo{volume}{105}},
  \bibinfo{pages}{177002} (\bibinfo{year}{2010}).

\bibitem[{\citenamefont{Mourik et~al.}(2012)\citenamefont{Mourik, Zuo, Frolov,
  Plissard, Bakkers, and Kouwenhoven}}]{Mourik25052012}
\bibinfo{author}{\bibfnamefont{V.}~\bibnamefont{Mourik}},
  \bibinfo{author}{\bibfnamefont{K.}~\bibnamefont{Zuo}},
  \bibinfo{author}{\bibfnamefont{S.~M.} \bibnamefont{Frolov}},
  \bibinfo{author}{\bibfnamefont{S.~R.} \bibnamefont{Plissard}},
  \bibinfo{author}{\bibfnamefont{E.~P. A.~M.} \bibnamefont{Bakkers}},
  \bibnamefont{and} \bibinfo{author}{\bibfnamefont{L.~P.}
  \bibnamefont{Kouwenhoven}}, \bibinfo{journal}{Science}
  \textbf{\bibinfo{volume}{336}}, \bibinfo{pages}{1003} (\bibinfo{year}{2012}).

\bibitem[{\citenamefont{Law et~al.}(2009)\citenamefont{Law, Lee, and
  Ng}}]{LawMaj}
\bibinfo{author}{\bibfnamefont{K.~T.} \bibnamefont{Law}},
  \bibinfo{author}{\bibfnamefont{P.~A.} \bibnamefont{Lee}}, \bibnamefont{and}
  \bibinfo{author}{\bibfnamefont{T.~K.} \bibnamefont{Ng}},
  \bibinfo{journal}{Phys. Rev. Lett.} \textbf{\bibinfo{volume}{103}},
  \bibinfo{pages}{237001} (\bibinfo{year}{2009}).

\bibitem[{\citenamefont{Flensberg}(2010)}]{Flencond}
\bibinfo{author}{\bibfnamefont{K.}~\bibnamefont{Flensberg}},
  \bibinfo{journal}{Phys. Rev. B} \textbf{\bibinfo{volume}{82}},
  \bibinfo{pages}{180516} (\bibinfo{year}{2010}).

\bibitem[{\citenamefont{Sau et~al.}(2010{\natexlab{b}})\citenamefont{Sau,
  Tewari, Lutchyn, Stanescu, and Das~Sarma}}]{Saucond}
\bibinfo{author}{\bibfnamefont{J.~D.} \bibnamefont{Sau}},
  \bibinfo{author}{\bibfnamefont{S.}~\bibnamefont{Tewari}},
  \bibinfo{author}{\bibfnamefont{R.~M.} \bibnamefont{Lutchyn}},
  \bibinfo{author}{\bibfnamefont{T.~D.} \bibnamefont{Stanescu}},
  \bibnamefont{and}
  \bibinfo{author}{\bibfnamefont{S.}~\bibnamefont{Das~Sarma}},
  \bibinfo{journal}{Phys. Rev. B} \textbf{\bibinfo{volume}{82}},
  \bibinfo{pages}{214509} (\bibinfo{year}{2010}{\natexlab{b}}).

\bibitem[{\citenamefont{Wimmer et~al.}(2011)\citenamefont{Wimmer, Akhmerov,
  Dahlhaus, and Beenakker}}]{wimmer2011quantum}
\bibinfo{author}{\bibfnamefont{M.}~\bibnamefont{Wimmer}},
  \bibinfo{author}{\bibfnamefont{A.~R.} \bibnamefont{Akhmerov}},
  \bibinfo{author}{\bibfnamefont{J.~P.} \bibnamefont{Dahlhaus}},
  \bibnamefont{and} \bibinfo{author}{\bibfnamefont{C.~W.~J.}
  \bibnamefont{Beenakker}}, \bibinfo{journal}{New J. Phys.}
  \textbf{\bibinfo{volume}{13}}, \bibinfo{pages}{053016}
  (\bibinfo{year}{2011}).

\bibitem[{\citenamefont{Fu}(2010)}]{Futelep}
\bibinfo{author}{\bibfnamefont{L.}~\bibnamefont{Fu}}, \bibinfo{journal}{Phys.
  Rev. Lett.} \textbf{\bibinfo{volume}{104}}, \bibinfo{pages}{056402}
  (\bibinfo{year}{2010}).

\bibitem[{\citenamefont{Sau et~al.}(2012)\citenamefont{Sau, Tewari, and
  Das~Sarma}}]{Sauexpt}
\bibinfo{author}{\bibfnamefont{J.~D.} \bibnamefont{Sau}},
  \bibinfo{author}{\bibfnamefont{S.}~\bibnamefont{Tewari}}, \bibnamefont{and}
  \bibinfo{author}{\bibfnamefont{S.}~\bibnamefont{Das~Sarma}},
  \bibinfo{journal}{Phys. Rev. B} \textbf{\bibinfo{volume}{85}},
  \bibinfo{pages}{064512} (\bibinfo{year}{2012}).

\bibitem[{\citenamefont{Fabrizio and Gogolin}(1994)}]{FabGog}
\bibinfo{author}{\bibfnamefont{M.}~\bibnamefont{Fabrizio}} \bibnamefont{and}
  \bibinfo{author}{\bibfnamefont{A.~O.} \bibnamefont{Gogolin}},
  \bibinfo{journal}{Phys. Rev. B} \textbf{\bibinfo{volume}{50}},
  \bibinfo{pages}{17732} (\bibinfo{year}{1994}).

\bibitem[{\citenamefont{Sengupta and Kim}(1996)}]{SenKim96}
\bibinfo{author}{\bibfnamefont{A.}~\bibnamefont{Sengupta}} \bibnamefont{and}
  \bibinfo{author}{\bibfnamefont{Y.}~\bibnamefont{Kim}},
  \bibinfo{journal}{\prb} \textbf{\bibinfo{volume}{54}}, \bibinfo{pages}{14918}
  (\bibinfo{year}{1996}).

\bibitem[{\citenamefont{Affleck}(1990)}]{affleck1990current}
\bibinfo{author}{\bibfnamefont{I.}~\bibnamefont{Affleck}},
  \bibinfo{journal}{Nucl. Phys. B} \textbf{\bibinfo{volume}{336}},
  \bibinfo{pages}{517} (\bibinfo{year}{1990}).

\bibitem[{\citenamefont{Affleck and
  Ludwig}(1991{\natexlab{a}})}]{affleck1991kondo}
\bibinfo{author}{\bibfnamefont{I.}~\bibnamefont{Affleck}} \bibnamefont{and}
  \bibinfo{author}{\bibfnamefont{A.}~\bibnamefont{Ludwig}},
  \bibinfo{journal}{Nucl. Phys. B} \textbf{\bibinfo{volume}{352}},
  \bibinfo{pages}{849} (\bibinfo{year}{1991}{\natexlab{a}}).

\bibitem[{\citenamefont{Affleck and
  Ludwig}(1991{\natexlab{b}})}]{affleck1991critical}
\bibinfo{author}{\bibfnamefont{I.}~\bibnamefont{Affleck}} \bibnamefont{and}
  \bibinfo{author}{\bibfnamefont{A.}~\bibnamefont{Ludwig}},
  \bibinfo{journal}{Nucl. Phys. B} \textbf{\bibinfo{volume}{360}},
  \bibinfo{pages}{641} (\bibinfo{year}{1991}{\natexlab{b}}).

\bibitem[{app()}]{app}
\bibinfo{note}{See the Supplementary Material, where we summarize our CFT
  considerations.}

\bibitem[{\citenamefont{Fabrizio and Gogolin}(1995)}]{FabGog2c}
\bibinfo{author}{\bibfnamefont{M.}~\bibnamefont{Fabrizio}} \bibnamefont{and}
  \bibinfo{author}{\bibfnamefont{A.~O.} \bibnamefont{Gogolin}},
  \bibinfo{journal}{Phys. Rev. B} \textbf{\bibinfo{volume}{51}},
  \bibinfo{pages}{17827} (\bibinfo{year}{1995}).

\bibitem[{\citenamefont{Fiete et~al.}(2008)\citenamefont{Fiete, Bishara, and
  Nayak}}]{FieteNayak}
\bibinfo{author}{\bibfnamefont{G.~A.} \bibnamefont{Fiete}},
  \bibinfo{author}{\bibfnamefont{W.}~\bibnamefont{Bishara}}, \bibnamefont{and}
  \bibinfo{author}{\bibfnamefont{C.}~\bibnamefont{Nayak}},
  \bibinfo{journal}{Phys. Rev. Lett.} \textbf{\bibinfo{volume}{101}},
  \bibinfo{pages}{176801} (\bibinfo{year}{2008}).

\bibitem[{\citenamefont{Law et~al.}(2010)\citenamefont{Law, Seng, Lee, and
  Ng}}]{Law2c}
\bibinfo{author}{\bibfnamefont{K.~T.} \bibnamefont{Law}},
  \bibinfo{author}{\bibfnamefont{C.~Y.} \bibnamefont{Seng}},
  \bibinfo{author}{\bibfnamefont{P.~A.} \bibnamefont{Lee}}, \bibnamefont{and}
  \bibinfo{author}{\bibfnamefont{T.~K.} \bibnamefont{Ng}},
  \bibinfo{journal}{Phys. Rev. B} \textbf{\bibinfo{volume}{81}},
  \bibinfo{pages}{041305} (\bibinfo{year}{2010}).

\bibitem[{\citenamefont{Nayak et~al.}(1999)\citenamefont{Nayak, Fisher, Ludwig,
  and Lin}}]{Nayak99}
\bibinfo{author}{\bibfnamefont{C.}~\bibnamefont{Nayak}},
  \bibinfo{author}{\bibfnamefont{M.~P.~A.} \bibnamefont{Fisher}},
  \bibinfo{author}{\bibfnamefont{A.~W.~W.} \bibnamefont{Ludwig}},
  \bibnamefont{and} \bibinfo{author}{\bibfnamefont{H.~H.} \bibnamefont{Lin}},
  \bibinfo{journal}{Phys. Rev. B} \textbf{\bibinfo{volume}{59}},
  \bibinfo{pages}{15694} (\bibinfo{year}{1999}).

\bibitem[{\citenamefont{Oshikawa et~al.}(2006)\citenamefont{Oshikawa, Chamon,
  and Affleck}}]{Oshikawa06}
\bibinfo{author}{\bibfnamefont{M.}~\bibnamefont{Oshikawa}},
  \bibinfo{author}{\bibfnamefont{C.}~\bibnamefont{Chamon}}, \bibnamefont{and}
  \bibinfo{author}{\bibfnamefont{I.}~\bibnamefont{Affleck}},
  \bibinfo{journal}{J. Stat. Mech. Theor. Exp.}
  \textbf{\bibinfo{volume}{2006}}, \bibinfo{pages}{P02008}
  (\bibinfo{year}{2006}).

\bibitem[{\citenamefont{Zee}(2010)}]{zee2010quantum}
\bibinfo{author}{\bibfnamefont{A.}~\bibnamefont{Zee}},
  \emph{\bibinfo{title}{Quantum field theory in a nutshell}}
  (\bibinfo{publisher}{Princeton University Press}, \bibinfo{year}{2010}).

\bibitem[{\citenamefont{Fuchs and Schweigert}(2003)}]{fuchs2003symmetries}
\bibinfo{author}{\bibfnamefont{J.}~\bibnamefont{Fuchs}} \bibnamefont{and}
  \bibinfo{author}{\bibfnamefont{C.}~\bibnamefont{Schweigert}},
  \emph{\bibinfo{title}{Symmetries, Lie algebras and representations: A
  graduate course for physicists}} (\bibinfo{publisher}{Cambridge University
  Press}, \bibinfo{year}{2003}).

\bibitem[{\citenamefont{Emery and Kivelson}(1992)}]{EK}
\bibinfo{author}{\bibfnamefont{V.~J.} \bibnamefont{Emery}} \bibnamefont{and}
  \bibinfo{author}{\bibfnamefont{S.}~\bibnamefont{Kivelson}},
  \bibinfo{journal}{Phys. Rev. B} \textbf{\bibinfo{volume}{46}},
  \bibinfo{pages}{10812} (\bibinfo{year}{1992}).

\bibitem[{\citenamefont{Coleman et~al.}(1995)\citenamefont{Coleman, Ioffe, and
  Tsvelik}}]{ColemanKondo1}
\bibinfo{author}{\bibfnamefont{P.}~\bibnamefont{Coleman}},
  \bibinfo{author}{\bibfnamefont{L.~B.} \bibnamefont{Ioffe}}, \bibnamefont{and}
  \bibinfo{author}{\bibfnamefont{A.~M.} \bibnamefont{Tsvelik}},
  \bibinfo{journal}{Phys. Rev. B} \textbf{\bibinfo{volume}{52}},
  \bibinfo{pages}{6611} (\bibinfo{year}{1995}).

\bibitem[{\citenamefont{Coleman and Schofield}(1995)}]{ColemanKondo2}
\bibinfo{author}{\bibfnamefont{P.}~\bibnamefont{Coleman}} \bibnamefont{and}
  \bibinfo{author}{\bibfnamefont{A.~J.} \bibnamefont{Schofield}},
  \bibinfo{journal}{Phys. Rev. Lett.} \textbf{\bibinfo{volume}{75}},
  \bibinfo{pages}{2184} (\bibinfo{year}{1995}).

\bibitem[{\citenamefont{Maldacena and Ludwig}(1997)}]{maldacena1997majorana}
\bibinfo{author}{\bibfnamefont{J.~M.} \bibnamefont{Maldacena}}
  \bibnamefont{and} \bibinfo{author}{\bibfnamefont{A.~W.~W.}
  \bibnamefont{Ludwig}}, \bibinfo{journal}{Nuclear Physics B}
  \textbf{\bibinfo{volume}{506}}, \bibinfo{pages}{565} (\bibinfo{year}{1997}).

\bibitem[{\citenamefont{Rahmani et~al.}(2010)\citenamefont{Rahmani, Hou,
  Feiguin, Chamon, and Affleck}}]{Rahmani10}
\bibinfo{author}{\bibfnamefont{A.}~\bibnamefont{Rahmani}},
  \bibinfo{author}{\bibfnamefont{C.-Y.} \bibnamefont{Hou}},
  \bibinfo{author}{\bibfnamefont{A.}~\bibnamefont{Feiguin}},
  \bibinfo{author}{\bibfnamefont{C.}~\bibnamefont{Chamon}}, \bibnamefont{and}
  \bibinfo{author}{\bibfnamefont{I.}~\bibnamefont{Affleck}},
  \bibinfo{journal}{Phys. Rev. Lett.} \textbf{\bibinfo{volume}{105}},
  \bibinfo{pages}{226803} (\bibinfo{year}{2010}).

\bibitem[{\citenamefont{Ludwig and Affleck}(1991)}]{ludwig1991exact}
\bibinfo{author}{\bibfnamefont{A.}~\bibnamefont{Ludwig}} \bibnamefont{and}
  \bibinfo{author}{\bibfnamefont{I.}~\bibnamefont{Affleck}},
  \bibinfo{journal}{\prl} \textbf{\bibinfo{volume}{67}}, \bibinfo{pages}{3160}
  (\bibinfo{year}{1991}).

\end{thebibliography}
\end{document}